# Maxwell Tension Supports the Water Bridge

A. Widom[1], Y.N. Srivastava[2], J. Swain[1], S. Sivasubramanian[3]

[1]Physics Department, Northeastern University, Boston MA 02115, USA

[2]Physics Department & INFN, University of Perugia, Perugia IT

[3]NSF Center for High-rate Nanomanufacturing, Northeastern University, Boston MA 02115 USA

'These authors contributed equally to this work'

**A cylindrical "flexible cable" made up of pure fluid water can be experimentally spanned across a spatial gap[1,2,3] with cable endpoints fixed to the top edges of two glass beakers. The cable has been called a "water bridge" in close analogy to iron cables employed to build ordinary span bridges. A necessary condition for the construction of a water bridge is that a large electric field exists parallel to and located within the water cable. Presently, there is no accepted detailed theory which quantitatively explains the forces which hold up the bridge. Our purpose is to present such theory based on the Maxwell pressure tensor[4] induced by the electric field albeit within the condensed matter dielectric fluid cable[5,6].**

The electromagnetic field carries both energy and momentum, and due to such physical properties can exert pressure. The nature of the pressure tensor forces was very clearly explained by Maxwell[4] in his treatise on the theory of electricity and magnetism. For an isotropic medium, be it the vacuum or a fluid dielectric[5,6] such as pure water, there are *pressure forces* normal to the directions of the electric field lines and *tension forces* parallel to the directions of the electric field lines. As an example of the Maxwell pressure force, one sees in Fig.1 a schematic drawing of parallel capacitor plates submerged into a dielectric fluid of mass density $\rho$ and dielectric constant $\varepsilon$. The fluid will rise between the capacitor plates[7] until the height $h$ wherein the Maxwell pressure force upwards would balance the gravitational force downward.

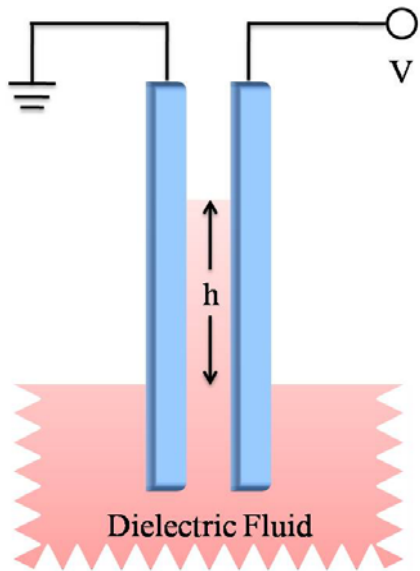

Fig.1: When parallel capacitor plates are submerged into a dielectric fluid, the Maxwell upward pressure force due to the electric field **E** between the plates must balance the gravitational acceleration **g** force downward on that liquid which has risen to a height $h = (\varepsilon - 1)E^2/(8\pi\rho g)$ wherein $\rho$ is the liquid mass density and $\varepsilon$ is the liquid dielectric constant

**Theoretical Method for the Maxwell Tension**

To theoretically explain the water bridge one may examine the experimental picture[2] of the suspended fluid cable as shown below in Fig.2.

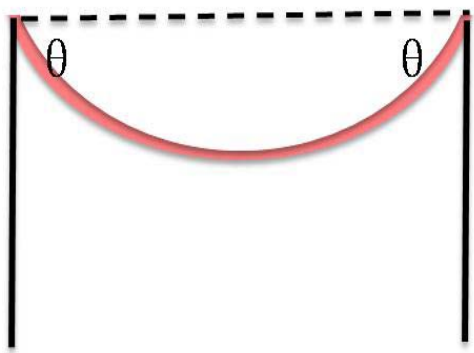

Fig.2: The water bridge consists of a flexible fluid cable which can be suspended by its endpoints. There is a sleight sag in the cable as befits the equilibrium of the total gravitational force *Mg* downward and the total Maxwell tension force $2\tau\sin\theta$ upward; i.e. *Mg* = $2\tau\sin\theta$ wherein $\theta$ is the angle between the cable tangent at the support and the horizontal.

The tension to weight ratio in the fluid cable of length $L$ in a water bridge is given via the Maxwell electric field pressure tensor as

$$\frac{\tau}{Mg} = \left(\frac{\varepsilon-1}{4\pi}\right)\frac{E^2}{\rho g L} \equiv \frac{\chi E^2}{\rho g L} = \frac{1}{2\sin\theta} \tag{1}$$

wherein $\tau$ is the flexible cable tension. If the tension is large compared with the weight, then the angle will be small and the sag in the water bridge will only be sleight.

As an approximate numerical experimental example[2,3] for a bridge with $L$=2.5 cm, $\varepsilon$=80, $\rho$=1 gm/cm$^3$, $g$=980 cm/sec$^2$, and $E$ =10 kilovolt/cm=33 gauss, one finds from Eq.(1) the suitably small angle $\theta$=10 °. This yields only a very slight sagging of the flexible cable.

While the above considerations concerning the Maxwell tension are thermodynamic in nature, a microscopic view is easily comprehensible. The fluid cable has an electric dipole moment per unit volume **P**=$\chi$**E** each of which has a negative charge localized near one end and an equal and opposite positive charge localized near the other end. With a large electric field within the cable, the dipole moments are aligned yielding an attraction between the plus side on one dipole moment and the negative side of the next dipole moment. The resulting force thereby yields a cable tension which supports the water bridge against gravity.

All correspondence should be sent to Yogendra.Srivastava@pg.infn.it